\documentclass{edm_article}
\usepackage{xcolor}
\usepackage[colorlinks=true,allcolors=blue]{hyperref}

\begin{document}

\title{Predicting Engagement in Video Lectures}
%
\numberofauthors{1}
\author{
Sahan Bulathwela, Mar\'ia P\'erez-Ortiz, Aldo Lipani, Emine Yilmaz and John Shawe-Taylor\\
       \affaddr{University College London, United Kingdom}\\
       \email{m.bulathwela@ucl.ac.uk}
}

\maketitle

\begin{abstract}
The explosion of Open Educational Resources (OERs) in the recent years creates the demand for scalable, automatic approaches to process and evaluate OERs, with the end goal of identifying and recommending the most suitable educational materials for learners. We focus on building models to find the characteristics and features involved in context-agnostic engagement (i.e. population-based), a seldom researched topic compared to other contextualised and personalised approaches that focus more on individual learner engagement. Learner engagement, is arguably a more reliable measure than popularity/number of views, is more abundant than user ratings and has also been shown to be a crucial component in achieving learning outcomes. In this work, we explore the idea of building a predictive model for population-based engagement in education. We introduce a novel, large dataset of video lectures for predicting context-agnostic engagement and propose both cross-modal and modality specific feature sets to achieve this task. We further test different strategies for quantifying learner engagement signals. We demonstrate the use of our approach in the case of data scarcity. Additionally, we perform a sensitivity analysis of the best performing model, which shows promising performance and can be easily integrated into an educational recommender system for OERs.
\end{abstract}

\category{K.3.1}{Computer Uses in Education}{Distance learning}
\category{H.3.1}{Content Analysis and Indexing}{Linguistic processing}

\terms{Human Factors, Measurement, Management}

\keywords{Context-free Engagement, Cold Start, Video lectures, Quality Assurance, Open Education, OER, Personalisation}

\section{Introduction}\label{sec:intro}

With the recent popularity of online learning platforms, the creation of Open Educational Resources (OERs) is increasing rapidly \cite{TVET2018}. 
This recent large-scale creation of educational material demands for ways to automatically manage educational resources. 
In the context of OERs, this means finding and recommending material that fits the learners' goals while maximising learning outcomes. Such a goal usually entails a large personalisation factor \cite{truelearn}.
We define it as \emph{contextualised engagement}, which captures how engaging a learning resource is with regard to the context of the learner (e.g., learning needs/goals and learner state). Although contextualised engagement has gained interest in the recent years \cite{sum_20}, we argue that there is also a \emph{context-agnostic engagement} factor, that only relates to features of the learning resource and attempts to capture the gold-standard label of population-based engagement (i.e. the marginal of contextual engagement for a resource across the population of learners). Modelling context-agnostic engagement enables identifying highly engaging resources across a population of learners before personalising educational recommendations to individuals.  This paper studies the features involved in context-agnostic engagement, as a first step towards building an integrative educative recommendation system, that will join both contextualised and context-agnostic features \cite{trueeducation}. 

A high quality learning resource needs to satisfy three main properties:
 i) academic soundness and appropriate coverage of the body of knowledge, 
ii) pedagogical robustness and 
iii) enabling learners to achieve their desired learning outcomes \cite{Lane10}. 
{Learner engagement} has been shown to be a proxy for (iii), as engaging with material is a prerequisite for learning. There is 
evidence from both online \cite{ramesh2014learning,lan2017behavior} and classroom \cite{pardos2014affective,slater2016semantic} educational settings showing that higher learner engagement increases the  likelihood of better learning outcomes. We thus focus on finding the general characteristics of engaging material. Using features that can be extracted across multiple modalities (video, text, audio etc.) allows developing prediction models for gold-standard engagement that are easily adaptable to a wide range of OERs and can be automated \cite{x5gon}.

Our work is one of the first to address educational engagement prediction with video lectures, specially from a quantitative perspective. 
One of our primary goals is to understand if easily automatable cross-modal features 
can be used as predictors for how engaging an educational resource is, as opposed to modality specific features. Although large-scale studies (involving millions of videos) have been conducted to analyse the prediction of engagement for general purpose videos \cite{beyondviews}, the largest study in the context of educational video lectures involves 800 videos from 4 courses and analyses engagement from a qualitative perspective \cite{Guo_vid_prod}. To the best of our knowledge, this work is the first attempt to predict engagement with educational videos automatically. Our experiments involve more than 4000 video lectures that span over 20 diverse subjects, making it the largest dataset to date in this field. Our dataset, code and best performing model are released with the paper. 

Given the usefulness of \emph{predicting context-agnostic engagement} and the scarcity of work in this topic, we are motivated to answer the following research questions, which will enable the deployment of such a model in an educational platform:
\begin{itemize}
    \item[RQ1] How to encode context-agnostic engagement? 
    \item[RQ2] How effective are cross-modal language-based features for predicting engagement with video lectures?
    \item[RQ3] Does including modality-specific features lead to a significant improvement in performance?
    \item[RQ4] What features influence context-agnostic engagement?
    \item[RQ5] Is predicting marginal population-based engagement useful over personalised engagement? 
    \item[RQ6] Can we assume a common underlying model for predicting engagement across different knowledge areas?
\end{itemize}

%

\section{Related Work}
\label{sec:related_work}

The interest in identifying useful and engaging information goes beyond the educational domain and is investigated in numerous other fields \cite{quality_features}. For example, Wikipedia uses a review system to evaluate the quality of its articles. To do so, different machine learning models, such as support vector regression and ensemble methods, are used 
with features such as text style, readability, structure, network, recency and review information \cite{Dalip_wiki_svr,wiki_wang}. Moreover, in the context of automatic essay scoring,  promising results have been obtained through rank preference support vector machines \cite{AES_dataset} and more sophisticated deep learning models \cite{taghipour2016neural}.

Quality-biased document ranking \cite{Bendersky2011} and spam web-page detection \cite{Ntoulas2006} are other areas in the information retrieval domain that also utilises textual features and recency related features.  
These features categorise into different verticals such as
understandability, topic coverage, presentation, freshness and authority \cite{quality_features}. 

OERs available to the public come in large-scale and various modalities \cite{x5gon,stud_eng_ou}, which makes modality-specific models of limited use. As existing work proposes models with domain/modality specific features (e.g. network features of Wikipedia \cite{dalip_quality_features} or speaker speed in videos \cite{Guo_vid_prod}), there is a need for models that can evaluate how engaging educational materials are at scale using a cross-modal feature set. We attempt to address this gap through this work.

\subsection{Why Modelling Engagement?}

As argued by Lane \cite{Lane10}, a well designed learning resource should enable the learner to achieve the expected learning outcomes. Prior work has studied \emph{learner engagement} in Massively Open Online Courses and shown that when optimised, \textbf{engagement can increase the likelihood of achieving better learning outcomes} \cite{ramesh2014learning,lan2017behavior}. User engagement has also been shown to differ greatly from popularity measures such as number of views \cite{beyondviews}, as the latter does not necessarily capture whether learners consume the material.
In our work, we also show that engagement does not positively correlate with user ratings. Instead, what we observe is that lectures with low rating also present low engagement rate. However, lectures with greater ratings can have different engagement rates. 

For videos, \emph{watch time} has been used as the main measure for quantifying engagement in the literature, e.g., for YouTube recommendations \cite{Covington2016}, predicting engagement with videos \cite{beyondviews}. For educational content, the median of normalised engagement time (i.e., the percentage of watch time from the total video)
 has been used as gold standard for engagement \cite{Guo_vid_prod}. Our work tests several approaches to encoding user engagement.

Most of the related work regarding predicting educational engagement attempts to model learner engagement as a function of the learner's context (demography, user activity, etc.) \cite{bonafini2017much,stud_eng_ou,beal2006classifying}, as opposed to modelling context-agnostic learner engagement as a function of content-based features of the educational resource, which is our aim. 
Context-agnostic engagement has been previously studied for video lectures, advocating for qualitative and general recommendations such as keeping videos short \cite{Guo_vid_prod}, using conversational language for lecture delivery \cite{brame2016effective} and others.
These recommendations empower authors to create better educational videos. However, none of these works address the need for automatically identifying the features of highly engaging educational resources, which is imperative for retrieving and recommending educational material at scale. 

\section{Data and Methodology}
\label{sec:methodology}

This section first describes the dataset built for predicting engagement, together with the set of features proposed in this paper. Then, we introduce the machine learning methods and the feature importance analysis method considered.

To address the research questions outlined in the introductory section of this paper we do the following: 
i) We 
study different ways of refining user engagement signals, linking to literature on psychometrics (RQ1).
ii) We 
propose two sets of easily automatable features for predicting engagement (cross-modal features inspired by context-agnostic quality literature and video-specific features) and evaluate the difference of predictive performance between them (RQ2 and RQ3).
iii) We
construct
a large dataset of video lectures 
and evaluate the performance of the proposed engagement signals and sets of features (RQ2-4).
iv) We 
compare cross-modal to modality specific features, analysing the impact of individual features in the predictive model that presents the most promising performance (RQ4).
v) 
We compare our population-based engagement approach to its personalised analogue to demonstrate its usefulness (RQ5). 
vi) We 
compare the engagement models obtained from dividing the video lectures in two differentiated knowledge areas: STEM (such as technology, physics and mathematics lectures) vs others (such as arts, social science and philosophy lectures). 


\subsection{Dataset and Features (RQ2-4)}\label{topic:data}
%
We use data from a popular OER repository, VideoLectures.Net (VLN)\footnote{\url{www.videolectures.net}}, a collection of videos of researchers presenting in peer-reviewed conferences. 
This data is suitable for our aim for two reasons: 
i) It contains watch patterns about how learners consume lectures, and 
ii) the lectures are peer-reviewed and hence material is controlled for correctness of knowledge and pedagogical robustness. The transcriptions of English lectures and English translations for the non-English lectures are provided by the TransLectures project\footnote{\url{www.translectures.eu}}. We restrict the final dataset to lectures that has been viewed by at least 5 unique users, leading to the final dataset having 4,063 lectures. These lectures are categorised into 21 subjects, e.g.~Computer Science, Physics, Philosophy, etc. Learner engagement labels of the dataset is computed using 155,850 user view log events (video viewing events) created between December 8, 2016 and February 17, 2018.The dataset constructed is publicly available, including different statistics of population engagement and all the cross-modal and video-based features proposed.

\subsubsection{Cross-modal Features}
\label{topic:features}

We selected a subset of cross-modal and mostly language-based features that are easy to extract from the VLN dataset.
The 13 extracted features are shown in Table \ref{table:features}.
This set has been selected based on recurring features in the related work  \cite{Bendersky2011,Dalip_wiki_svr,Guo_vid_prod,Ntoulas2006,wiki_wang} and their quality verticals \cite{quality_features} identified in our prior work. 
The majority of features were extracted using methods and token (word) sets that are found in the prior work referenced in Table \ref{table:features}. 

Additionally, we introduce the \emph{published date}, represented by converting the video publication date to UNIX epoch time (in days). In other words, it is the number of days between January 01, 1970 and the lecture published date. 



\subsubsection{Video-based Features}

We also extracted four out of the seven features proposed for analysing educational engagement with video lectures from \cite{Guo_vid_prod}, selecting those features that can be automatised and are objective. These are: 
i) \emph{lecture duration}, as shorter videos have been shown to be much more engaging; 
ii) \emph{is chunked}, whether the lecture has been partitioned into multiple parts;
iii) a set of indicator variables describing the \emph{type of lecture}, such as tutorial, workshop, etc; and 
iv) \emph{speaker speed}, measured by the average amount of words spoken per minute. We also include the \emph{silence period rate (SPR)}, calculated using the special tags in the video transcripts that indicate silence. Formally, for a lecture $\ell$, this feature $\texttt{SPR}(\ell)$ is calculated as follows:
\begin{equation}\label{eq:SPR}
    \texttt{SPR}(\ell) =
    \frac{1}{D(\ell)}
    \sum_{t \in T(\ell)}
    D(t) \cdot \mathcal{I}(N(t) = \texttt{"silence"}),
\end{equation}
where $t$ is a tag in the collection of tags $T(\ell)$ that belong to lecture $\ell$, $N$ returns the type of tag $t$ and $D$ returns the duration of tag $t$ or lecture $\ell$ and $\mathcal{I}(\cdot)$ is the indicator function (returning 1 when the condition is verified, 0 otherwise).

%
%


\begin{table}
\caption{Extracted features from the VLN dataset.}
\label{table:features}
\centering
\begin{tabular}{ |l|c|} \hline
    Feature&Reference\\ \hline
    \multicolumn{2}{|c|}{\emph{Content-based features}} \\ \hline
    Easiness (FK Easiness)& \cite{Dalip_wiki_svr} \\ \hline 
    Stop-word Presence Rate& \cite{Ntoulas2006} \\ \hline
    Stop-word Coverage Rate&  \cite{Ntoulas2006} \\ \hline

    Document Entropy& \cite{Bendersky2011} \\ \hline
    Word Count& \cite{wiki_wang}  \\ \hline
    Title Word Count& \cite{Bendersky2011} \\ \hline

    Preposition Rate& \cite{Dalip_wiki_svr} \\ \hline
    Auxiliary Rate& \cite{Dalip_wiki_svr} \\ \hline
    To Be Rate& \cite{Dalip_wiki_svr} \\ \hline
    Conjunction Rate& \cite{Dalip_wiki_svr} \\ \hline
    Normalization Rate& \cite{Dalip_wiki_svr} \\ \hline
    Pronoun Rate& \cite{Dalip_wiki_svr} \\ \hline
        Published Date& --- \\ \hline
    
    \multicolumn{2}{|c|}{\emph{Video-based features}} \\ \hline    
    
    Lecture Duration& \cite{Guo_vid_prod} \\ \hline
    Is Chunked& \cite{Guo_vid_prod} \\ \hline
    Video Lecture Type& \cite{Guo_vid_prod} \\ \hline
    Speaker speed& \cite{Guo_vid_prod} \\ \hline
        Silence Period Rate (SPR)& --- \\ \hline

\end{tabular}
\end{table}

\subsection{Quantifying Engagement (RQ1)}\label{topic:labels}

Our work focuses on implicit user feedback (most specifically, engagement). Implicit feedback (in the form of number of views, engagement or any other  measure that does not require the user to provide explicit feedback) has been used for building recommender systems for nearly two decades with great success \cite{Oard1998ImplicitFF,Jannach2018RecommendingBO,Jawaheer14}, as an alternative to explicit ratings, which have a high cognitive load on users and thus are usually sparse. However, implicit signals have other challenges associated with them. For example, implicit feedback is usually positive-only \cite{Jannach2018RecommendingBO} and can contain effects such as popularity bias, i.e., there might be a bias towards more popular items, whereas implicit feedback for other items may be very sparse. 
There has been several works  investigating the relationship between explicit and implicit
feedback \cite{Claypool:2001:III:359784.359836,ShapiraTM06,Zigoris2006}, which we also do through this work. 

The main measure that we use to quantify engagement is the \textbf{Median of Normalised Engagement/watch Time (MNET)}, as it has been proposed as the gold standard for engagement with educational materials in previous work \cite{Guo_vid_prod}.
To have the MNET label in the range $[0,1]$, we set the upper bound of MNET to 1. 
We observed in our initial data analysis that MNET values 
 in the VLN dataset 
follow a Log-Normal distribution, 
where it can be seen that most users generally abandon the lecture after a generally low time threshold. We hypothesise this may be because it takes some time to decide whether the content is relevant for the learner. Users that make it after this threshold seem more committed and thus the leaving rate is significantly lower. 
To address this, as this is usually a problem when using machine learning methods, we applied a log transformation to transform the engagement signal.
The final label, \emph{Log Median Normalised Engagement Time (LMNET)} is computed using the following:
%
\begin{equation}
\label{eq:engagement_rate}
    \texttt{LMNET}(\ell) = \ln (\max(\texttt{MNET}(\ell), 1)).
\end{equation}


To test if LMNET can be further improved,
we compare this approach of encoding engagement to other alternative ways of quantifying and cleaning engagement signals, drawing inspiration from the literature on psychometrics and subjective assessment \cite{Janowski15,Thurstone1927}, which focuses on explicit human feedback and assumes that users present cognitive biases and differences, with applications in preference ranking and measuring perception-based qualities, such as engagement. The intuition behind this is that different learners may have a different engagement threshold and scale, similarly as with explicit ratings \cite{Janowski15}. We compare different approaches for defining engagement:
\begin{enumerate}
    \item \textbf{Raw LMNET}, as per Eq. (\ref{eq:engagement_rate}) which considers that no user differences exist and the marginal over the population can be directly used as gold standard label for engagement, similarly as in \cite{Guo_vid_prod}.
    \item \textbf{Cleaned LMNET}, for which we test the removal of bot-like users (those users with an average engagement rate less than 5\%), which may have a detrimental factor in the median of raw engagement.
    \item \textbf{Standardised LMNET}, in which we preprocess LMNET per user (subtracting the mean of the user and dividing by the standard deviation), as commonly done with human ratings in order to remove user biases and differences \cite{Janowski15}. In this scale, positive values indicate lectures that are more engaging than the mean of the user and vice versa.
    \item \textbf{Comparative MNET}, in which we exploit the law of comparative judgement and use psychometric scaling to go from user comparative engagement data to a probabilistically interpretable engagement scale \cite{Thurstone1927, perezortiz2017practical}. More specifically, we assume that engagement data can only be compared per user (as users may have different biases, thresholds or engagement scales). To do so, we generated a matrix of engagement comparisons (of the type: Did learner $i$ prefer lecture $A$ to $B$ in terms of engagement?), which is used as the input for psychometric scaling, producing a final scale in which distances can be interpreted in terms of probability of greater engagement.
\end{enumerate} 
As discussed, the limitation of these approaches is that they disregard the context of the learner and the temporal component that may inherently be present when engaging with educational material.  A different measure to encode engagement is found in Wu et al. \cite{beyondviews}, where the main idea is to compare engagement relative to the length of the video. The authors propose this for entertainment videos. However, we argue against this approach in the case of educational material, as the aim is to take the learner to the desired state in the most efficient way, thus the general recommendations found in the literature of keeping videos as short as possible \cite{Guo_vid_prod}.

\subsection{Machine Learning Models (RQ2)} \label{topic:ml_eval}


To learn to rank video lectures based on engagement, we evaluate the performance using pointwise ranking models. Regression algorithms predict the target variable in real value space ($y \in \mathbb{R}$), which allows them to create a global ranking of observations based on predictions. We also evaluate the performance of engagement prediction using kernelised models. Kernelisation allows capturing non-linear patterns in data without having to operate in the respective basis. Although it is more computationally efficient than working in the non-linear space itself, it is more computationally expensive than solving the non-kernelised problem. Our choice of kernel for the models is the Radial Basis Function (RBF). RBF  kernel is widely used in the literature and has mathematical connections to other popular kernels such as exponential and polynomial kernels  \cite{chang2010training,shawe2004kernel}.  

We use two regression algorithms, namely, \emph{Ridge Regression (RR)} and \emph{Support Vector Regression (SVR)} in primal form. We use RR as it is a widely used  algorithm for regression \cite{beyondviews} and SVR as it has performed well in a similar task in prior work \cite{Dalip_wiki_svr}. We also evaluate the performance of the kernelised version of the same two algorithms (with RBF kernel), \emph{Kernelised Ridge Regression (KRR)} and \emph{Kernelised Support Vector Regression (KSVR)}. This allows us to understand if there is non-linearity in the patterns that benefits the prediction task. In all four models discussed above, we employ standard scaling as these models are not scale invariant. L2 regularisation is used to defend against overfitting and multicollinearity \cite{Ng_l2}. As ensemble techniques have shown to perform well in prior work \cite{wiki_wang}, we also employ a \emph{Random Forest Regressor (RF)} to evaluate its prediction capabilities. This model is also capable of capturing non-linear patterns.



\subsubsection{Comparison to Personalised Models (RQ5)}
\label{topic:use_of_context_free}

One of our aims is to compare the population-based model to its personalised counterpart. The idea in this case is to test if a common baseline can be assumed for all users. For this, we train the same machine learning models per user, using the features previously proposed.


\subsection{Feature Importance Analysis (RQ4)}
\label{topic:feat_imp}

Understanding how different features influence engageability of materials is vital in educational domain as learners will be guided on life-changing pathways based on these judgements. In a conventional linear model such as RR or SVM, feature importance analysis is straightforward as the weight coefficients reflect the influence of features. 

In this paper we use \emph{SHapley Additive exPlanations (SHAP)}, which is a model-agnostic framework that quantifies the impact of features on the model predictions. It reliably estimates feature importance of complex model families such as ensembles \cite{NIPS2017_shap}.
A SHAP value is computed for every feature of every prediction. Given a prediction and a feature, 
SHAP is computed by averaging how the prediction changes when the feature is present and vice versa. This procedure enables quantifying the contribution of each feature to the model prediction.
By plotting all the SHAP values of the prediction data points in a SHAP summary plot, we can identify how each feature influences the prediction. By calculating the \emph{Mean Absolute SHAP (MAS)} for each feature $f$ over the observations: 
\begin{equation}\label{eq:maSv}
\text{MAS}_f = 
\frac{1}{N}
\sum_{n=1}^{N}\left|\ \text{SHAP}_{f, n}\ \right|,
\end{equation}
we obtain a more quantitative understanding of feature influence. $N$ is the number of observations.


\section{Experiments and Discussion}
\label{sec:results}
This section shows the experimental setup and results for the different experiments conducted.

\subsection{Experimental Setup}
\label{topic:exp_setup}

The evaluation of the machine learning models is performed using a 5-fold cross-validation for both feature sets.    
The performance of different machine learning models with different engagement quantification approaches can be found in Table \ref{table:accuracy}. The performance when video-specific features are added is found in Table \ref{table:vid_spec_result}. 

After gaining an understanding of model performance (see results in Table \ref{table:accuracy}), we employ the best performing method and encoding for the rest of the analyses, using a hold-out validation with a train-test split of 70:30 to save computation. 
That is, the model is trained on the 70\% training set and interpreted using the 30\% test set. The experiments were implemented using \texttt{Scikit-learn} \cite{scikit-learn}, \texttt{textatistic} \cite{textatistic} 
and \texttt{SHAP} \cite{NIPS2017_shap} python packages. 
The source code in python and dataset are publicly available\footnote{\url{https://github.com/sahanbull/context-agnostic-engagement}}.

\subsubsection{Evaluation metrics}

\emph{Pairwise accuracy (Pair.)} and \emph{Spearman Rank Order Correlation Coefficient (SROCC)} are the ranking metrics we used to evaluate the ranking performance of machine learning models with different engagement signal encodings.

Identifying models that can rank between video lectures is the core objective of this work. Hence, we devise \emph{pairwise accuracy} as the main evaluation metric. Pairwise accuracy is more intuitive for this task as it represents the fraction of pairwise comparisons where the model could predict the more engaging lecture.
Another opportunity that pairwise comparison provides is the ability to restrict the comparisons to subsets of lecture pairs (e.g. lectures that belong to the same subject, lectures that have similar LMNET).

In some of our experiments we also perform misranking analysis and report the pairwise accuracy.
Misranking could happen if a subset of examples is systematically difficult to rank. We hypothesize that misclassification happens more frequently as the difference of LMNET between a pair of video lectures gets smaller. That is, the model may struggle to differentiate between two lectures with similar engagement. By doing this analysis, we can also understand the sensitivity of the prediction model to similarly engaging lectures. Obviously, misranking a pair of lectures that are significantly different in engagement incurs a larger cost in terms of user satisfaction than misranking a pair of lectures with similar engagement.

\subsubsection{Controlling for Topics in Content} 
\label{topic:content_control}
The topics covered in the content of the lecture is likely to drive learner engagement. For instance, Data Science lectures can be more popular than Physics lectures leading to easy pairwise comparison predictions between the domains. To test this, we restrict in some experiments the pairwise accuracy calculation to pairs of lectures that belong to the same domain (\emph{subject-specific} column in Table \ref{table:vid_spec_result}) and observe if the accuracy value changes significantly compared to its counterpart metric that considers all lecture pairs in a domain-agnostic fashion. 



\subsection{Results}
This section presents a series of experiments to: 
\begin{itemize}
    \item[E1] Analyse the relationship between engagement, number of views and mean star ratings (RQ1). 
    \item[E2] Test different machine learning models and engagement signals for the cross-modal features (RQ1-2).
    \item[E3] Study the distribution of engagement with respect to length of materials (RQ4).
    \item[E4] Study the influence of modality-specific features and comparison across subject areas (RQ3).
    \item[E5] Analyse the importance of different features in the model (RQ4).
    \item[E6] Compare the population-based model to its personalised counterpart (RQ5).
    \item[E7] Test if the same underlying model can be assumed for different knowledge/subject areas (RQ6).
\end{itemize}

\begin{figure*}[ht!]
  \centering
  \includegraphics[width=.95\textwidth]{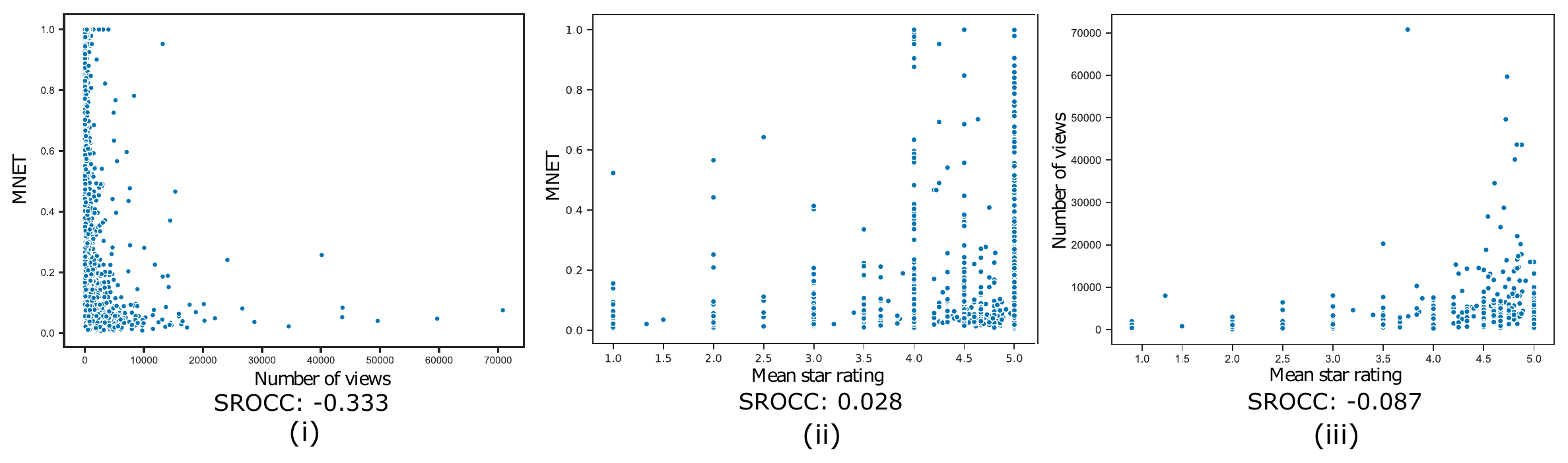}
  \caption{Scatter plots showing the relationship between (i) number of views vs. MNET, (ii) mean star rating for the video lecture vs. MNET and (iii) mean star rating vs. number of views, together with the Spearman's rank correlation coefficient (SROCC).}
  \label{fig:ratings}
\end{figure*}

\subsubsection{E1: Engagement vs Views and Ratings}

The VLN data source also has mean star ratings (explicit feedback) for a subset of the considered lectures. It is noteworthy that we only have access to mean star ratings, not to the individual ratings per observer or the number of measurements. As done in previous work, we also analyse the relationship between implicit signals (engagement and number of views) and explicit ratings. This can be found in Figure { \ref{fig:ratings}}, where we show mean star rating vs MNET and number of views. The SROCC is close to zero, mainly because of the large number of lectures with high rating but low engagement and number of views. We test the correlation for the 4 different versions of engagement considered (raw, cleaned, standardised and comparative), but all achieve similar results, with SROCC close to zero. One conclusion that is clear from the plot in Figure { \ref{fig:ratings}} is that number of views, ratings and engagement do represent very different information. For example, it can be appreciated that the variance of MNET and number of views increases with higher ratings, showing heteroskedasticity. This indicates that for low quality resources (with low ratings) engagement is generally low, whereas for resources with higher ratings engagement differs and may be either high or low. This suggest other factors involved in engagement than simply quality perceived by learners. Regarding number of views it seems that the correlation is rather negative, showing that the materials with the highest number of views present very low engagement.

\begin{table*}[t]
\small
 \setlength{\tabcolsep}{3pt}
\caption{Pairwise accuracy (Pair.) and Spearman's Rank Correlation Coefficient(SROCC) of engagement prediction models with standard error from 5-fold cross validation and cross-modal features. }\label{table:accuracy}
\begin{tabular}{|c|l|l|l|l|l|l|l|l|l|l|}
\hline
Model & \multicolumn{2}{c|}{RR} & \multicolumn{2}{c|}{SVR} & \multicolumn{2}{c|}{KRR} & \multicolumn{2}{c|}{KSVR} & \multicolumn{2}{c|}{RF} \\ \hline
Engagement & \multicolumn{1}{c|}{Pair.} & \multicolumn{1}{c|}{SROCC} & \multicolumn{1}{c|}{Pair.} & \multicolumn{1}{c|}{SROCC} & \multicolumn{1}{c|}{Pair.} & \multicolumn{1}{c|}{SROCC} & \multicolumn{1}{c|}{Pair.} & \multicolumn{1}{c|}{SROCC} & \multicolumn{1}{c|}{Pair.} & \multicolumn{1}{c|}{SROCC} \\ \hline
Raw & .705$\pm$.011 & .581$\pm$.027 & .707$\pm$.000 & .586$\pm$.000 & \textit{.715$\pm$.004} & \textit{.607$\pm$.011} & .714$\pm$.007 & .604$\pm$.019 & \textbf{.723$\pm$.009} & \textbf{.625$\pm$.027} \\ \hline
Clearned & .636$\pm$.033 & .396$\pm$.093 & .634$\pm$.031 & .392$\pm$.089 & .646$\pm$.025 & .424$\pm$.071 & .642$\pm$.028 & .414$\pm$.078 & .646$\pm$.031 & .427$\pm$.087 \\ \hline
Standard & .603$\pm$.035 & .302$\pm$.098 & .600$\pm$.035 & .292$\pm$.100 & .609$\pm$.035 & .315$\pm$.099 & .602$\pm$.025 & .297$\pm$.071 & .611$\pm$.035 & .323$\pm$.099 \\ \hline
Comparative & .624$\pm$.010 & .365$\pm$.028 & .624$\pm$.012 & .363$\pm$.036 & .626$\pm$.013 & .370$\pm$.040 & .627$\pm$.009 & .373$\pm$.027 & .636$\pm$.012 & .397$\pm$.038 \\ \hline
\end{tabular}
\end{table*}

\subsubsection{E2: Encoding and Predicting Engagement} 
Inherently, the task of finding a better engagement signal is very challenging, given the lack of ground truth. In this paper, we first attempt to see if any of these signals present better correlation with star ratings.
However, we observe from Figure { \ref{fig:ratings}} that engagement is not strongly correlated with perceived quality by users (explicit star ratings) and similar results emerge for different methods of quantifying engagement, meaning it is inconclusive that transforming raw engagement signals strengthens its relationship to explicit perceived quality. 
Thus, in order to decide on which is the best way of capturing and quantifying engagement, we compare the pairwise accuracy for the four proposed approaches (raw LMNET, cleaned, standardised and comparative). This simply tells us which output target variable is easier to predict given the proposed features. Table \ref{table:accuracy} presents these results, together with the pairwise accuracy (Pair.) and Spearman's Rank Order Correlation Coefficient (SROCC) obtained for each machine learning model with the standard error bounds based on 5-fold cross validation. The larger the accuracy value, the better performing the model is. 

These results suggest that raw LMNET may be the most appropriate target label, particularly since the proposed features seem to be more useful when building a model for predicting raw LMNET. These results do not contradict the literature, both educational and non-educational, as MNET has been used as the gold-standard way of quantifying engagement. Our experiments thus showed that the use of subjective assessment inspired transformations do not improve the predictive power of engagement signals. 
 This may be because these transformations/correction methods are initially designed to address biases in latent user preferences. 
 Although similar biases may exist in learners when consuming educational materials (e.g. learner fatigue, different engagement thresholds, language level preferences, etc.) we hypothesise that the most influential driver of engagement is the information content and style of the video. 

Another observation from Table \ref{table:accuracy} is that KRR and KSVR models outperform their linear versions. This suggests that there could be non-linearity in the dataset that is better captured by the kernel techniques. RF seems to be the best performing model providing more evidence that non-linearity plays a significant role.

\begin{figure}[t!]
  \centering
  \includegraphics[width=0.9\linewidth]{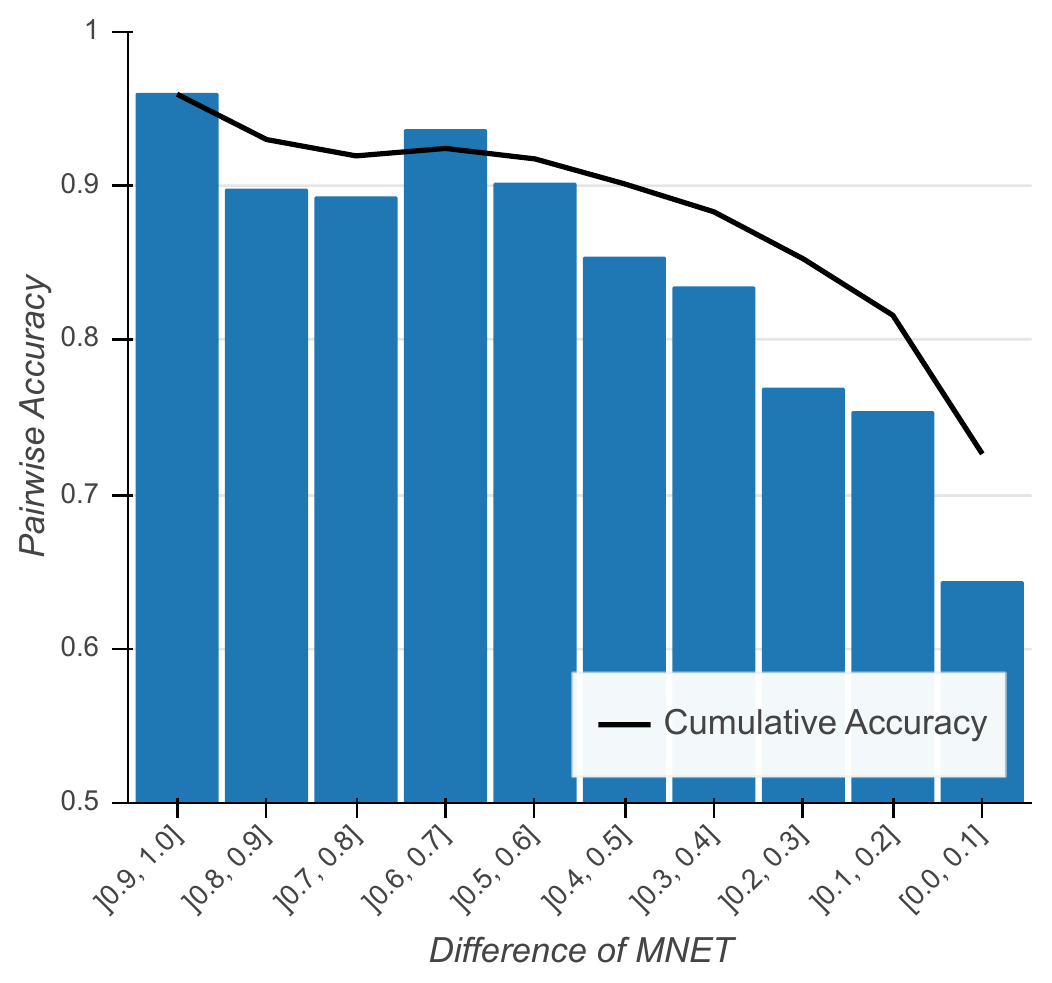}
  \caption{Bar chart plot showing how the pairwise accuracy changes based on the difference of  MNET between lecture pairs }
  \label{fig:misclassifications}
\end{figure}

To show how the accuracy changes when the difference of MNET between two lectures changes, we first compute all the possible differences between pairs of lectures and binarize these pairs into bins of size $0.1$ from $0$ to $1$, finally we compute the pairwise accuracy for each bin.
Figure \ref{fig:misclassifications} shows how the performance of the model changes based on the \emph{difference of MNET between lecture pairs}. The bars in the figure represent the pairwise accuracy for all the pairs that belong to the same bin. For example, the pairs with largest difference of MNET are predicted correctly with 0.962 accuracy whereas pairs with the smallest difference are predicted with 0.642 accuracy. 

Intuitively, a learner might have a similar experience consuming a pair of video lectures that are similarly engaging (at least disregarding the topic), as one is less likely to notice the difference. The black line in Figure \ref{fig:misclassifications} presents the \emph{cumulative pairwise accuracy} of the model if we were to assume that the learners are insensitive to noticing the difference of experience for lecture pairs that have a small difference of MNET. The plotted cumulative pairwise accuracy (y-axis) is computed by restricting the comparisons to lecture pairs with a difference of MNET between the lower bound of the x-axis value and 1.0.  
For instance, the cumulative pairwise accuracy of the model is 0.816 when the learners do not notice the difference when interacting with similarly engaging lecture pairs with MNET difference of [0.0, 0.2]. This value is the pairwise accuracy of all the lecture pairs with a MNET difference of ]0.2, 1.0].  

\subsubsection{E3: Length of Materials vs. Engagement}
\label{topic:length_comp}

\begin{figure}[ht!]
  \centering
  \includegraphics[width=0.9\linewidth]{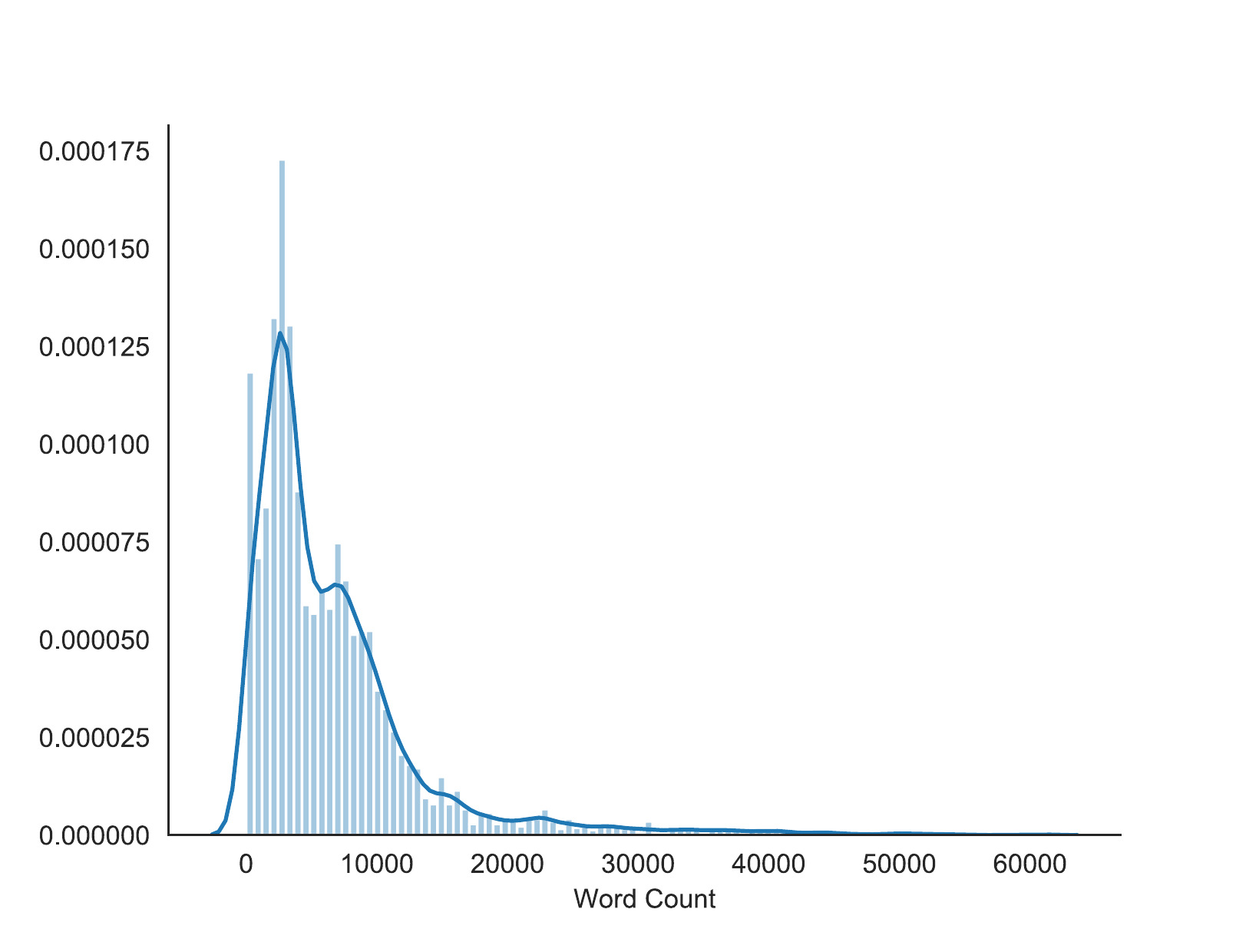}
  \caption{Distribution of word count of video lectures}
  \label{fig:word_count_dist}
\end{figure}

\begin{figure}[h]
\centering
  \includegraphics[width=0.9\linewidth]{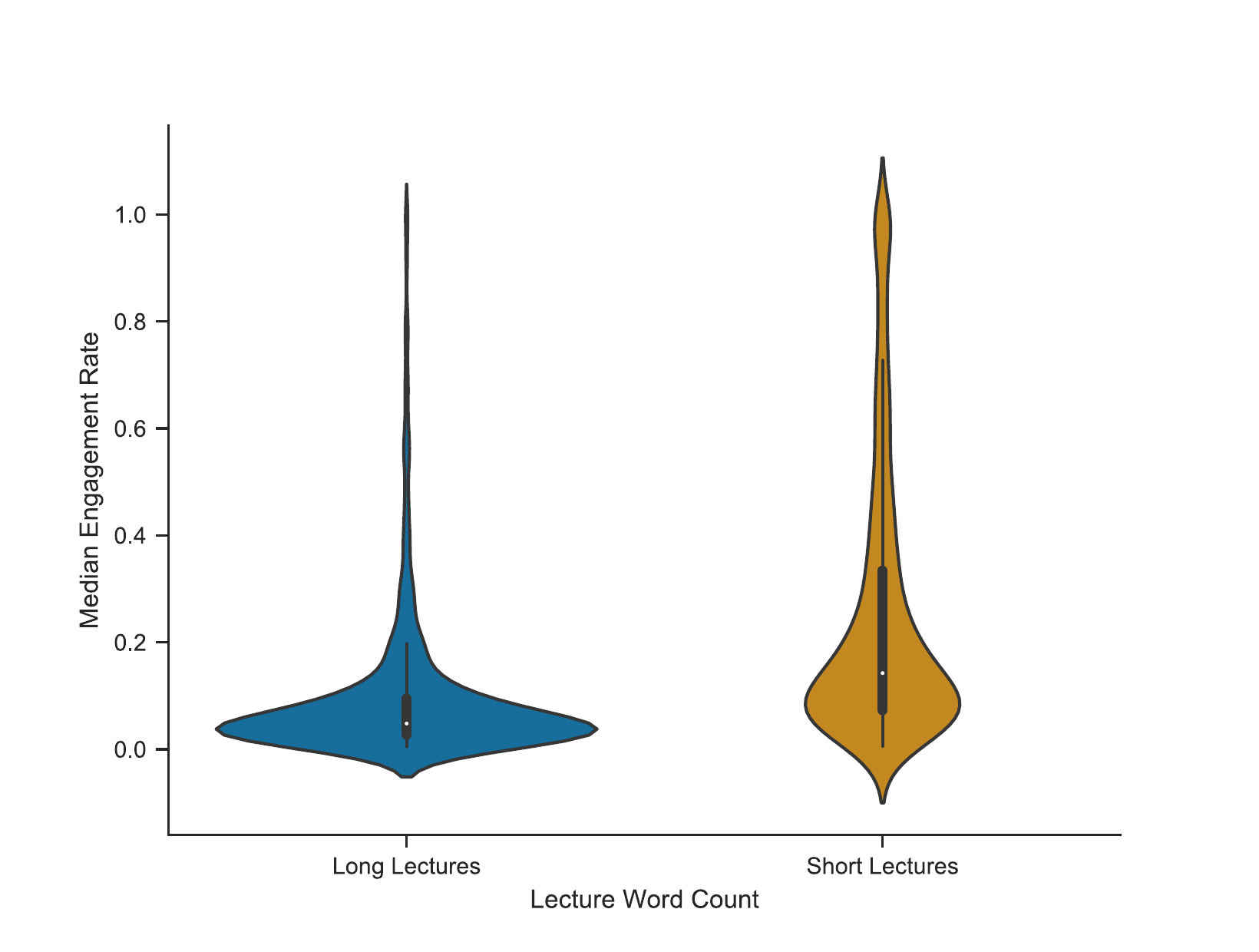}
  \caption{Distribution of engagement labels for short and long lectures.}
  \label{fig:violin_plot}
\end{figure}

Several studies have shown that features that quantify material length have a significant impact (this is also reaffirmed by our observations in our feature importance analysis in Figure \ref{fig:shap} and \ref{fig:shap_vid}) on sustained engagement with the material \cite{Guo_vid_prod,Dalip_wiki_svr}. We investigate how the length of the lectures impacts engagement prediction (i.e. if the engagement predictor is na\"ively distinguishing between long vs. short video lectures). We first investigate the distribution of total word count in the video lectures (Figure \ref{fig:word_count_dist}), which is directly related to the length. Based on the observed multi-modal distribution, we make two groups, i) short lectures of less than 5000 words and ii) long lecture (see engagement distribution in Figure \ref{fig:violin_plot}). It can be seen that, as anticipated, the percentage of watch time tends to be shorter for long lectures.

\begin{figure}[h]
  \centering
  \includegraphics[width=0.95\linewidth]{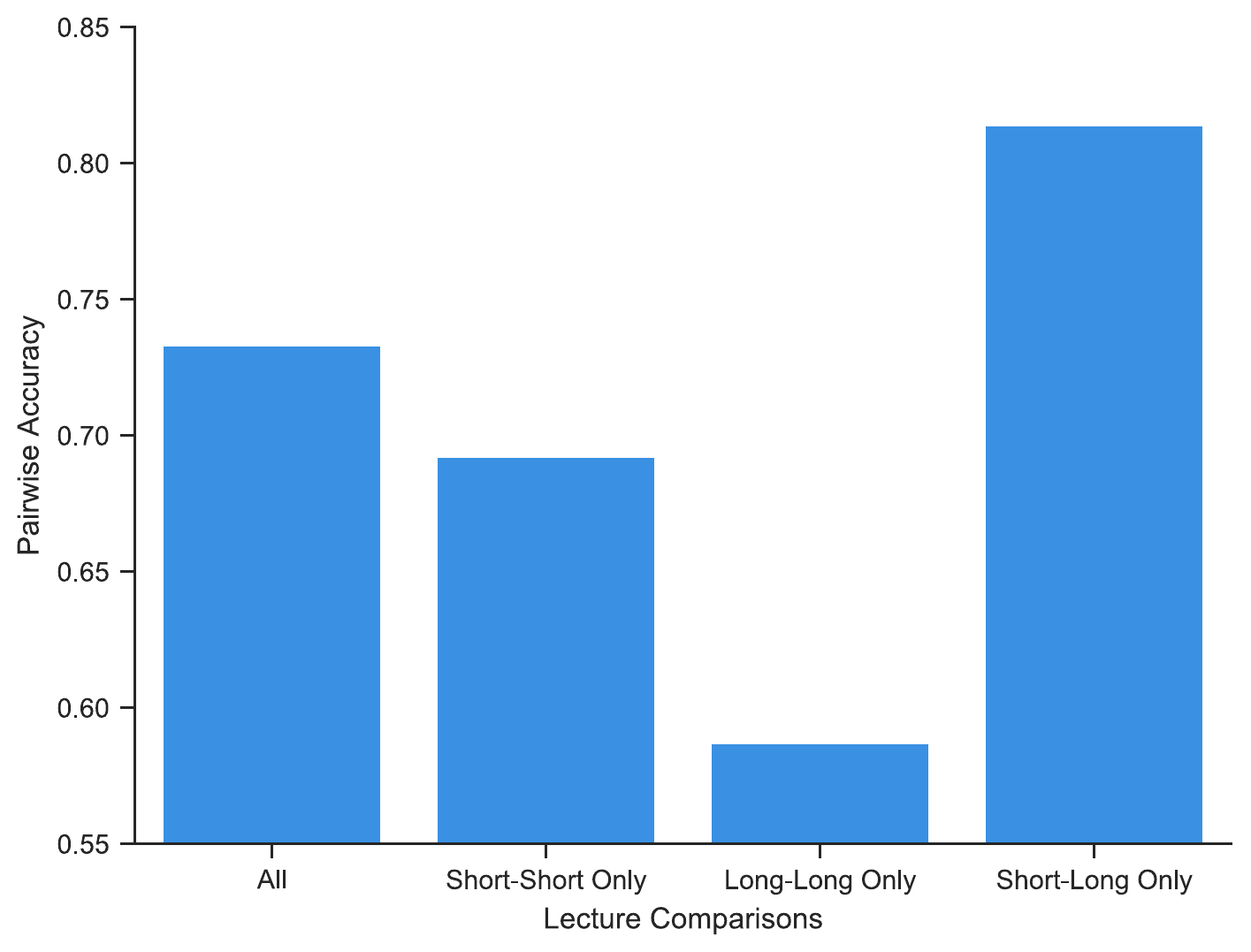}
  \caption{Accuracy bar chart for different types of comparisons using short and long lecture labels.}
  \label{fig:lengthwise_accuracy}
\end{figure}

 We investigate how median engagement labels are distributed in the aforementioned groups and also how the pairwise accuracy differs among and between the groups. 
Figure \ref{fig:lengthwise_accuracy} shows that the model is better at comparing between short-short lecture pairs compared to long-long lecture pairs. In the context of VLN dataset, this is good because there are more short lectures than long lectures (Figure \ref{fig:word_count_dist}). Recent findings (e.g.\cite{Guo_vid_prod}) also encourage authors to make short videos, increasing the likelihood of future video productions being short lectures. MNET distribution in Figure \ref{fig:violin_plot} shows that long lectures have a more skewed target value distribution concentrated closer to 0 compared to short lectures suggesting that learners tend to consume smaller fractions of long videos. This is likely to be driven by factors beyond other measured features of the lectures, such as limited time availability and short attention span of learners.


\begin{table}[!t]
    \small
    \centering
    \caption{Pairwise accuracy with standard error via 5-fold cross validation for RF model using content-based features vs. content-based + video-specific features.}\label{table:vid_spec_result}
    \begin{tabular}{ |l|c|c| } 
    \hline
    Model & \multicolumn{2}{c|}{Pairwise Accuracy}\\ 
    & Subject-agnostic & Subject-specific \\ \hline

    Content-based Features &.724$\pm$.014  &.733$\pm$.018\\  \hline
    Video-specific Features & \textbf{.744$\pm$.011}  & \textbf{.755$\pm$.014} \\  \hline
    \end{tabular}
        \label{tab:video}
\end{table}

\subsubsection{E4: Video-Features and Subject Areas}
Table \ref{tab:video} shows how the pairwise accuracy increases when restricted to subject-specific comparisons (lecture pairs belonging to the same subject area). This is clearly an advantage, given that most often, an educational recommendation system needs to make choices among sets of resources that belong to the same subject area. 


Table \ref{tab:video} additionally shows how the performance differs when using exclusively the cross-modal set of features and when adding video specific features. The addition of video features increase the performance by approximately 2\%. This result shows that there is a compromise in performance when restricting features to cross-modal features although the feature extractors can be reused in a practical scenario.

\subsubsection{E5: Feature Importance Analysis}
\begin{figure}[t]
\includegraphics[width=0.98\columnwidth]{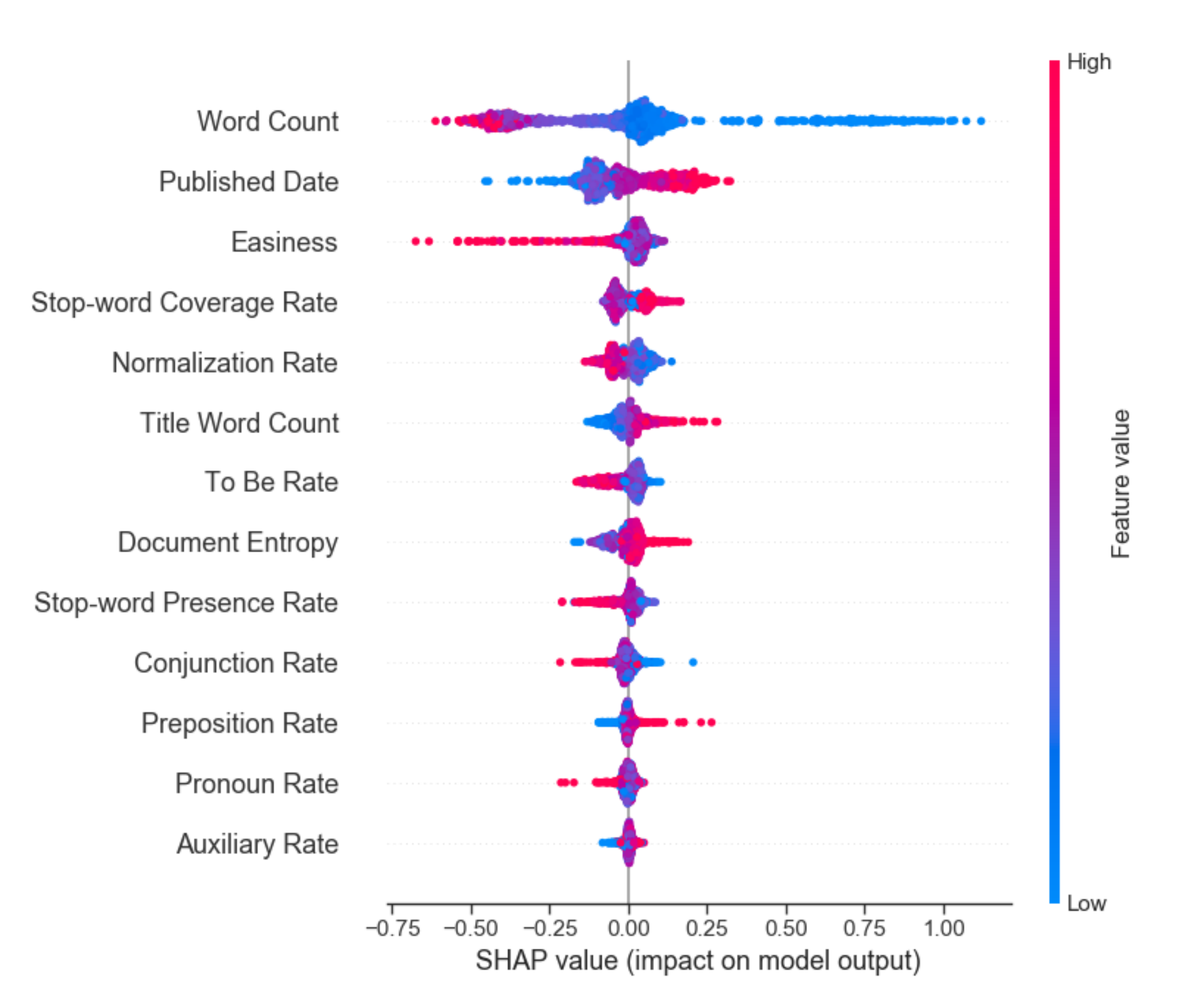}
\caption{SHAP summary plot for cross-modal features.}
\label{fig:shap}
\end{figure}

\begin{figure}[t]
\includegraphics[width=0.98\columnwidth]{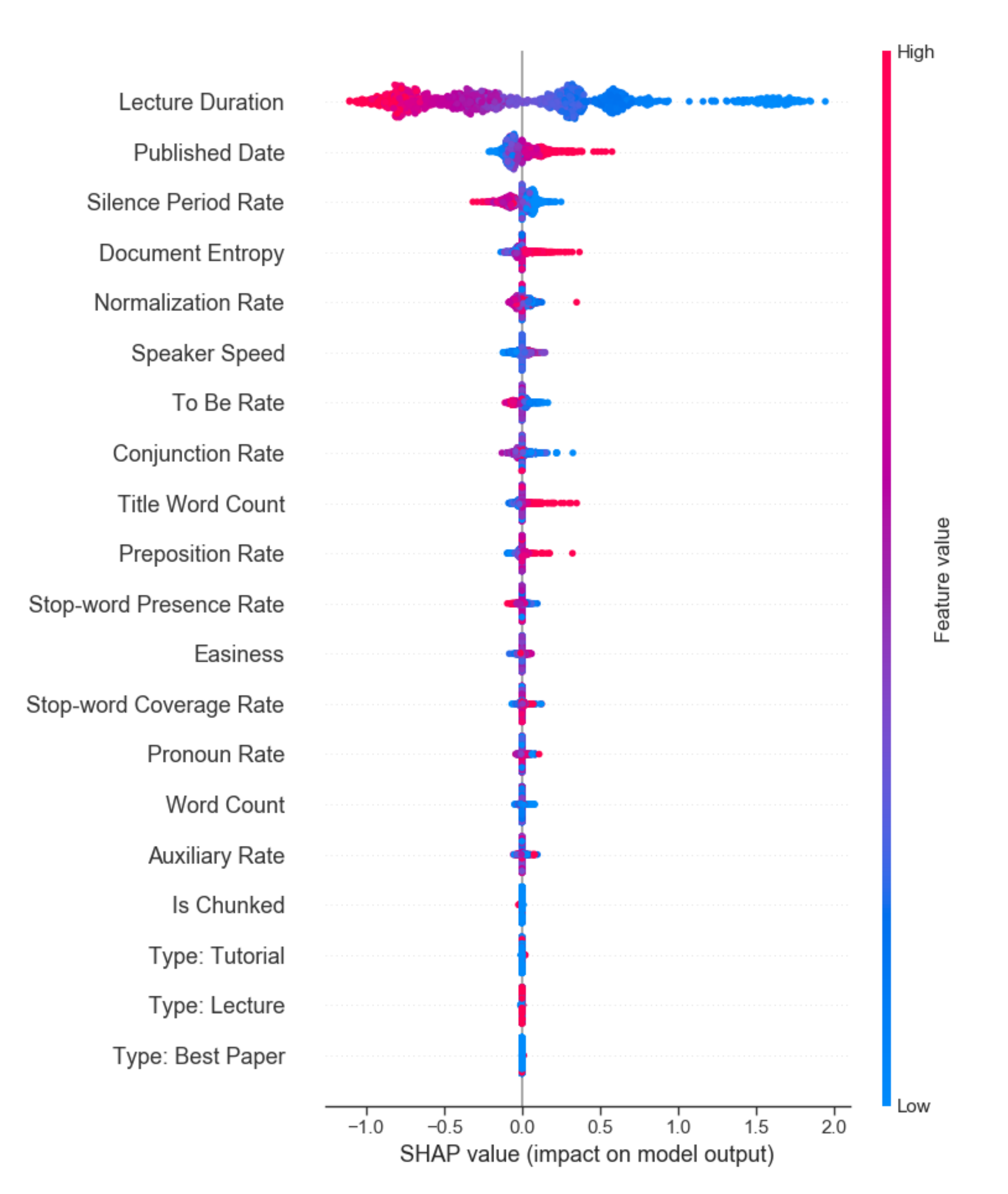}
\caption{SHAP summary plot with video-specific features.}
\label{fig:shap_vid}
\end{figure}

The SHAP value summary plots for content-based and video-specific feature sets are presented in Figures \ref{fig:shap} and \ref{fig:shap_vid} respectively, where the features are ordered based on overall feature influence using the best performing prediction model (RF).
Colour represents the raw feature value (blue low, red high). For example, when the observed values of a feature is red and they have a negative SHAP value, this means that higher values of this feature negatively impact LMNET prediction. 
Regarding video length, figures validate its impact on engagement, showing that long videos generally present lower engagement and vice versa, with lecture duration and word count being the most relevant features. Prior studies confirm this observation \cite{Guo_vid_prod,beyondviews,dalip_quality_features}).

\begin{table}[!t]
 \setlength{\tabcolsep}{3pt}
\small
\caption{Influence of content-based features on engagement as per their verticals outlined in \cite{quality_features}.}
\label{tab:feat_imp}
\begin{tabular}{ |l|l|r|r|} 
    \hline
    Quality Vertical&Feature&MAS&\% MAS\\ \hline
    Topic Coverage & Word Count & .250 & .366 \\ \hline
    Freshness & Published Date & .107 & .157 \\ \hline
    Understandability & Easiness & .052 & .076 \\ \hline
    Understandability & Stop-word Coverage Rate & .042 & .061 \\ \hline
    Presentation & Normalization Rate & .039 & .058 \\ \hline
    Topic Coverage & Title Word Count & .039 & .057 \\ \hline
    Presentation & To Be Rate & .038 & .055 \\ \hline
    Topic Coverage & Document Entropy & .033 & .048 \\ \hline
    Understandability & Stop-word Presence Rate & .028 & .041 \\ \hline
    Presentation & Conjunction Rate & .019 & .028 \\ \hline
    Presentation & Preposition Rate & .014 & .020 \\ \hline
    Presentation & Pronoun Rate & .013 & .020 \\ \hline
    Presentation & Auxiliary Rate & .009 & .013\\ \hline
\end{tabular}
\end{table}

Table \ref{tab:feat_imp} complements Figure \ref{fig:shap} by giving a more quantitative representation of how the influence of different features across the test dataset changes. 
Higher MAS is associated with more important features.
 By looking at the five most influential features, we observe that all identified quality verticals (topic coverage, understandability, freshness and presentation) are represented. This observation supports the importance of considering all the different verticals 
 when predicting context-agnostic engagement. 
The influence of top features is also consistent with results on quality biased information search \cite{Bendersky2011} where it is also found that \textit{Title Word Count} is comparatively less important. Figures \ref{fig:shap} and \ref{fig:shap_vid} also show the importance of modality-specific features in this prediction task by raising \emph{Lecture Duration}, \emph{Silence Period Rate} and \emph{Speaker Speed} in Figure \ref{fig:shap_vid} to high ranks.

\subsubsection{E6: Population-based vs. Personalised}
\begin{figure}[ht!]
  \centering
  \includegraphics[width=.93\linewidth]{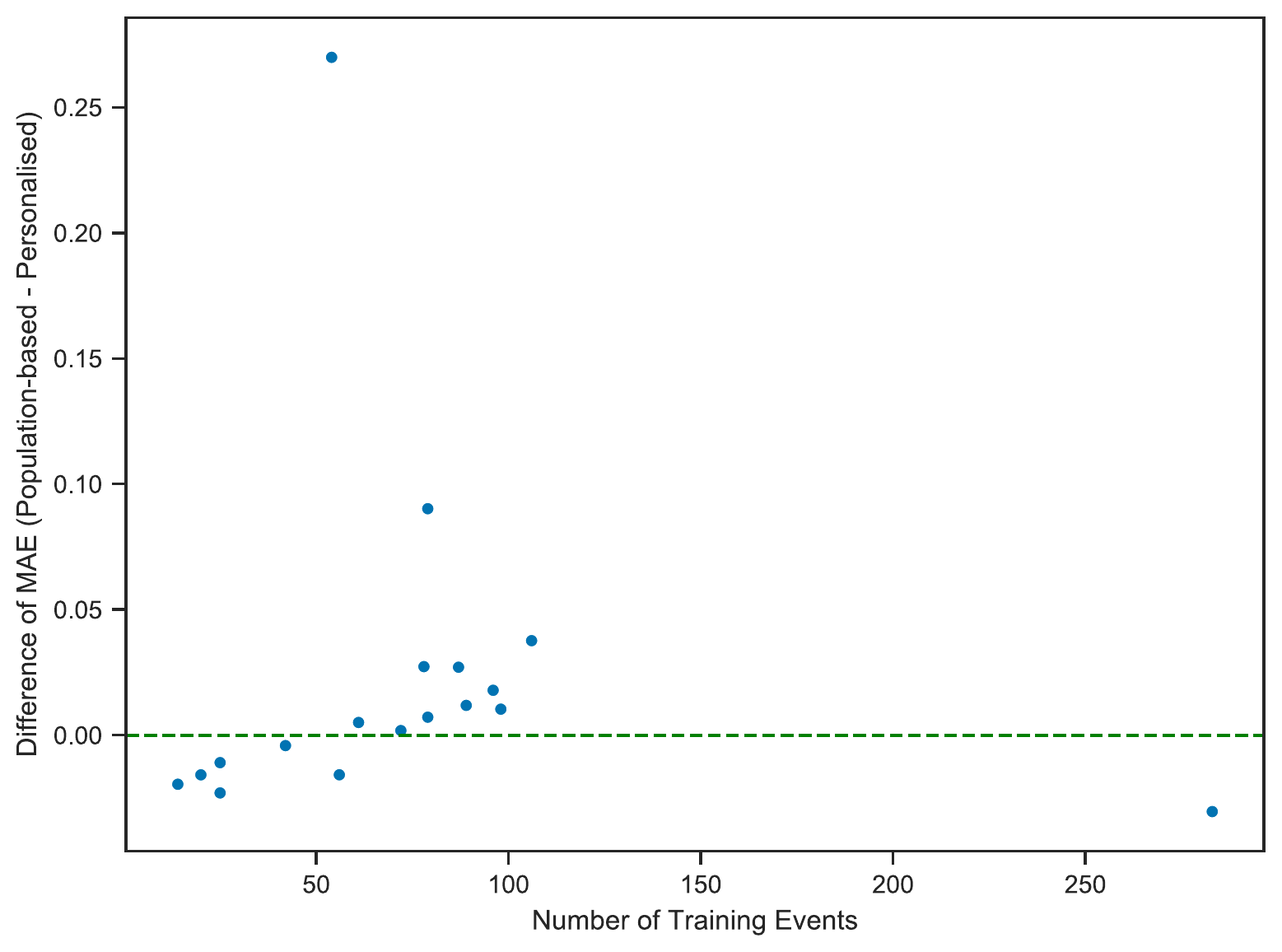}
  \caption{How the difference between Mean Absolute Error (MAE) of population-based and personalised models change with the number of training events per learner. 
  Each data point is an individual learner in the dataset.}
  \label{fig:margin_person}
\end{figure}

We use the 20 most active learners from the VLN dataset to compare the predictive performance of context-agnostic to contextual/personalised models when predicting engagement. {Firstly, we train the population-based prediction model using the VLN dataset (outlined in section \ref{topic:data}) using a 70:30 train-test split.
} In order to build the personalised model, for each user, we make a {similar 70:30} train-test split respecting the temporal order of their individual events. 
We use the training data to build a personalised model per user using {only} the cross-modal set of features {(no video-specific features)}.
For each learner $\ell$, we make predictions on the $N_\ell$ test events using (i) population-based model 
and (ii) the personalised model trained on personal events of the learner. We calculate Mean Absolute Error ($MAE(\ell)$) as:
\begin{equation}\label{eq:mae}
\text{MAE}(\ell) = 
\frac{1}{N_\ell}
\sum_{n=1}^{N_\ell}\left|\ y_n - \hat{y}_n \right|,
\end{equation}
where $\hat{y}_n$ is the prediction.
As regression models are devised for the task, $MAE$ is a sensible evaluation metric to measure predictive performance of the models.
Then we calculate the difference of $MAE(\ell)$ between the population-based and personalised model. Thus, a negative value indicates that the population model is better and vice versa. 
Figure \ref{fig:margin_person}, where the y-axis represents the difference in performance between the population-based and personalised, shows that the population-based model has better predictive power when the number of training examples available for the individual learner is limited ($\approx 60$). This is represented by the green line (at a MAE difference of 0). This demonstrates the usefulness of the population-based engagement prediction model in a situation where the recommender system is in a cold-start phase. 

\subsubsection{E7: Individual Models per Knowledge Area}

To understand if training subject-specific models can improve on the predictive power of the overall task, we partition the lecture records into 2 categories:
\begin{itemize}
    \item \textbf{STEM:} \textit{Life Sciences, Physics, Technology and Mathematics}.
    \item \textbf{Miscellaneous:} \textit{Social Sciences, Humanities, Arts and Philosophy}.
\end{itemize}
Then, we compare the performance of the models trained on subject-agnostic (STEM + miscellaneous) and subject-specific (STEM only or miscellaneous only) training data. Table \ref{tab:subject} demonstrates that there is little evidence in our results contradicting that a common subject-agnostic engagement model can be assumed across knowledge areas. This is shown in the fact that both training with all knowledge areas or dividing into two, the models obtain very similar test accuracy for each category (.737 vs .732 and .708 vs .704). In fact, the best performance is obtained in both cases by training with the whole dataset. This indicates that in general a common engagement model can be assumed throughout knowledge areas.

\begin{table}[!t]
    \small
    \centering
    \caption{Pairwise accuracy for STEM and Miscelleneous (Misc.) lectures when trained with subject-agnostic and subject-specific training data}
    \begin{tabular}{ |l|c|c| } 
    \hline
    Training Data  & \multicolumn{2}{c|}{Test Data}\\ 
    & STEM & Misc. \\ \hline

    Subject-agnostic  & .737   & .708\\  \hline
    Subject-specific & .732 & .704 \\  \hline

    \end{tabular}
        \label{tab:subject}
\end{table}

\subsection{Limitations}

Firstly,
the model does not include features that capture authority of content or its authors. 
Authority has been identified as an influential feature and lacking it is a weakness of this model. However, identifying an authority indicator that generalises beyond niche communities (e.g. academia) is challenging yet necessary, especially in the OER landscape where anyone can author learning materials. Additionally, the topic coverage features used in this model (\textit{Word Count}, \textit{Title Word Count} and \textit{Document Entropy}) are relatively na\"ive, although they are useful. Having better features will likely improve the model.
The current work demonstrates promise in predicting learner engagement with video lectures using easily automatable material features alone. More sophisticated features, both cross-modal and modality-specific could lead to higher predictive performance and better understanding of context-agnostic engagement.
Thirdly, the engagement model is trained on English lectures and English translation of non-English lectures. This impacts the generalisation ability of the model.  
The same applies to non-video content as well. More rigorous testing is needed in these fronts. Lastly, given that our dataset only considers OERs and excludes the learning dimension, we highlight that some of our findings may not be directly applicable to other type of educational material. Particularly, given that most of our features are language-based and we disregard visual information, the built models may not generalise to general purpose videos.



\section{Conclusions}
\label{sec:conclusion}

Given its timely need, we set out to develop and empirically test the suitability of engagement prediction models for automatically assessing context-agnostic engagement of OERs. 
Due to the scarcity of publicly available datasets for the task, we sourced a new video-lectures dataset and evaluated how different machine learning models perform on this dataset. 
In our analysis, we observed that the Random Forest algorithm performs best. 
We show that cross-modal features provide satisfactory performance, which is a major advantage, since these can be extracted from different resource modalities. Further experiments show that the predictive performance of the model can gain a slight boost in performance by adding modality-specific features. However, the performance does not deviate significantly.
Feature analysis showed that lecture length features are the most influential features in predicting context-agnostic engagement, which agrees with prior work. Other moderately influential features come from diverse quality verticals. Our analysis also showed that the model classifies much better when lectures with very different engagement values are compared, as opposed to lectures with similar engagement. This is natural and obviously the negative impact of misranking pairs of similar engagement lectures is relatively small. Our experiments demonstrated that the built model is useful in data scarcity scenarios, e.g. to approach the common cold-start problem in recommender systems. This is both for new users and new content, as our model can automatically estimate the engagement for new material and the model can be used as a prior for when we do not have enough data from a user to build a personalised model. We finally show that dividing the dataset into different knowledge areas (Subjects) and building separate models does not show improved performance, thus validating that a common underlying model can be built for estimating engagement across differentiated knowledge areas.

The proposed context-agnostic engagement prediction model can be beneficial in improving different components of an educational recommendation system. In situations where new content is discovered frequently (e.g. OER landscape \cite{x5gon,x5learn}), the proposed prediction model estimates \emph{how engaging materials are} prior to exposing them to the learner population. This allows better balancing the risks relating to learner satisfaction with opportunities of having fresh materials. Also, the proposed context-agnostic model can be integrated with a personalisation system in different ways. It can act as a prior that mitigates cold-start problem both on user and content fronts. In systems where personalisation heavily focuses on the topics covered in the materials \cite{trueeducation,truelearn}, this model can complement the content-based model by accounting for stylistic and lingual features that go beyond topic coverage.

To further improve the models, future work should address the three main limitations discussed: Future versions of our model should incorporate more sophisticated features. It could be beneficial to include features capturing \emph{authority} and \emph{topic coverage} \cite{quality_features}. In this sense, Wikification \cite{wikifier} can be used to extract covered topics, and data driven authority features, such as \cite{Adler_wiki_content_driven}, can be used to learn a universal author authority score.
  In the cross-modal front, more features focusing on content understanding, such as topic coherence and argument strength, can be considered. In the video-specific front, features such as liveliness of the presenter, sound quality and narration quality can be incorporated. 
Regarding the generalisation capabilities of the model, evaluating the effectiveness of the cross-modal feature set with a bigger video lecture dataset \cite{Guo_vid_prod,beyondviews} and a text dataset \cite{Dalip_wiki_svr} will increase the confidence on the feature set. Similarly, non-English datasets should also be taken into account.

\section{Acknowledgments}
This research is part of the X5GON project funded from the EU's Horizon 2020 research programme grant No 761758 and partially funded by the EPSRC Fellowship titled "Task Based Information Retrieval", under grant No EP/P024289/1.

%
\bibliographystyle{abbrv}
\bibliography{sigproc}  

\begin{thebibliography}{10}

\bibitem{Adler_wiki_content_driven}
B.~T. Adler and L.~de~Alfaro.
\newblock A content-driven reputation system for the wikipedia.
\newblock In {\em Proc.~of Int. Conf. on World Wide Web}, 2007.

\bibitem{beal2006classifying}
C.~R. Beal, L.~Qu, and H.~Lee.
\newblock Classifying learner engagement through integration of multiple data
  sources.
\newblock In {\em Proc.~of AAAI Conference on Artificial Intelligence}, 2006.

\bibitem{Bendersky2011}
M.~Bendersky, W.~B. Croft, and Y.~Diao.
\newblock Quality-biased ranking of web documents.
\newblock In {\em Proc.~of ACM Int. Conf. on Web Search and Data Mining}, 2011.

\bibitem{bonafini2017much}
F.~Bonafini, C.~Chae, E.~Park, and K.~Jablokow.
\newblock How much does student engagement with videos and forums in a mooc
  affect their achievement?
\newblock {\em Online Learning Journal}, 21(4), 2017.

\bibitem{brame2016effective}
C.~J. Brame.
\newblock Effective educational videos: Principles and guidelines for
  maximizing student learning from video content.
\newblock {\em CBE Life Sciences Education}, 15(4), 2016.

\bibitem{wikifier}
J.~Brank, G.~Leban, and M.~Grobelnik.
\newblock Annotating documents with relevant wikipedia concepts.
\newblock In {\em Proc.~of Slovenian KDD Conf. on Data Mining and Data
  Warehouses (SiKDD)}, 2017.

\bibitem{x5learn}
S.~Bulathwela, S.~Kreitmayer, and M.~P\'{e}rez-Ortiz.
\newblock What's in it for me? augmenting recommended learning resources with
  navigable annotations.
\newblock In {\em Proceedings of the 25th International Conference on
  Intelligent User Interfaces Companion}, IUI 20, 2020.

\bibitem{sum_20}
S.~Bulathwela, M.~P\'{e}rez-Ortiz, R.~Mehrotra, D.~Orlic, C.~de~la Higuera,
  J.~Shawe-Taylor, and E.~Yilmaz.
\newblock Sum20: State-based user modelling.
\newblock In {\em Proceedings of the 13th International Conference on Web
  Search and Data Mining}, WSDM 20, pages 899--900, New York, NY, USA, 2020.
  Association for Computing Machinery.

\bibitem{trueeducation}
S.~Bulathwela, M.~Perez-Ortiz, E.~Yilmaz, and J.~Shawe-Taylor.
\newblock Towards an integrative educational recommender for lifelong learners.
\newblock In {\em AAAI Conference on Artificial Intelligence}, AAAI 20, 2020.

\bibitem{truelearn}
S.~Bulathwela, M.~Perez-Ortiz, E.~Yilmaz, and J.~Shawe-Taylor.
\newblock Truelearn: A family of bayesian algorithms to match lifelong learners
  to open educational resources.
\newblock In {\em AAAI Conference on Artificial Intelligence}, AAAI 20, 2020.

\bibitem{quality_features}
S.~Bulathwela, E.~Yilmaz, and J.~Shawe-Taylor.
\newblock {Towards Automatic, Scalable Quality Assurance in Open Education}.
\newblock In {\em Workshop on AI and the United Nations SDGs at Int. Joint
  Conf. on Artificial Intelligence}, 2019.

\bibitem{chang2010training}
Y.-W. Chang, C.-J. Hsieh, K.-W. Chang, M.~Ringgaard, and C.-J. Lin.
\newblock Training and testing low-degree polynomial data mappings via linear
  svm.
\newblock {\em Journal of Machine Learning Research}, 11(Apr):1471--1490, 2010.

\bibitem{Claypool:2001:III:359784.359836}
M.~Claypool, P.~Le, M.~Wased, and D.~Brown.
\newblock Implicit interest indicators.
\newblock In {\em Proceedings of the 6th International Conference on
  Intelligent User Interfaces}, IUI '01, pages 33--40, New York, NY, USA, 2001.
  ACM.

\bibitem{Covington2016}
P.~Covington, J.~Adams, and E.~Sargin.
\newblock Deep neural networks for youtube recommendations.
\newblock In {\em Proc.~of ACM Conf. on Recommender Systems}, 2016.

\bibitem{dalip_quality_features}
D.~H. Dalip, M.~A. Goncalves, M.~Cristo, and P.~Calado.
\newblock A general multiview framework for assessing the quality of
  collaboratively created content on web 2.0.
\newblock {\em Journal of the Association for Information Science and
  Technology}, 2017.

\bibitem{Dalip_wiki_svr}
D.~H. Dalip, M.~A. Gon\c{c}alves, M.~Cristo, and P.~Calado.
\newblock Automatic assessment of document quality in web collaborative digital
  libraries.
\newblock {\em Journal of Data and Information Quality}, 2(3), Dec. 2011.

\bibitem{TVET2018}
M.~Ehlers, R.~Schuwer, and B.~Janssen.
\newblock Oer in tvet: Open educational resources for skills development, 2018.

\bibitem{Guo_vid_prod}
P.~J. Guo, J.~Kim, and R.~Rubin.
\newblock How video production affects student engagement: An empirical study
  of mooc videos.
\newblock In {\em Proc.~of the First ACM Conf. on Learning @ Scale}, 2014.

\bibitem{textatistic}
E.~Hengel.
\newblock {Publishing while Female. Are women held to higher standards?
  Evidence from peer review}.
\newblock Cambridge Working Papers in Economics 1753, 2017.

\bibitem{stud_eng_ou}
M.~Hussain, W.~Zhu, W.~Zhang, and S.~M.~R. Abidi.
\newblock Student engagement predictions in an e-learning system and their
  impact on student course assessment scores.
\newblock {\em Computational Intelligence and Neuroscience}, 2018(6347186),
  2018.

\bibitem{Jannach2018RecommendingBO}
D.~Jannach, L.~Lerche, and M.~Zanker.
\newblock Recommending based on implicit feedback.
\newblock In {\em Social Information Access}, 2018.

\bibitem{Janowski15}
L.~{Janowski} and M.~{Pinson}.
\newblock The accuracy of subjects in a quality experiment: A theoretical
  subject model.
\newblock {\em IEEE Transactions on Multimedia}, 17(12):2210--2224, Dec 2015.

\bibitem{Jawaheer14}
G.~Jawaheer, P.~Weller, and P.~Kostkova.
\newblock 8 modeling user preferences in recommender systems: A classification
  framework for explicit and implicit user feedback.
\newblock {\em ACM Transactions on Information and System Security}, 2:26
  pages, 05 2014.

\bibitem{lan2017behavior}
A.~S. Lan, C.~G. Brinton, T.-Y. Yang, and M.~Chiang.
\newblock Behavior-based latent variable model for learner engagement.
\newblock In {\em Proc.~of Int. Conf. on Educational Data Mining}, 2017.

\bibitem{Lane10}
A.~Lane.
\newblock Open information, open content, open source.
\newblock In {\em The Tower and The Cloud}, pages 158--168. 2010.

\bibitem{NIPS2017_shap}
S.~M. Lundberg and S.-I. Lee.
\newblock A unified approach to interpreting model predictions.
\newblock In {\em Advances in Neural Information Processing Systems}. 2017.

\bibitem{Ng_l2}
A.~Y. Ng.
\newblock Feature selection, l1 vs. l2 regularization, and rotational
  invariance.
\newblock In {\em Proc.~of Int. Conf. on Machine Learning}, 2004.

\bibitem{x5gon}
E.~Novak, J.~Urbancic, and M.~Jenko.
\newblock Preparing multi-modal data for natural language processing.
\newblock In {\em Proc.~of Slovenian KDD Conf. on Data Mining and Data
  Warehouses (SiKDD)}, 2018.

\bibitem{Ntoulas2006}
A.~Ntoulas, M.~Najork, M.~Manasse, and D.~Fetterly.
\newblock Detecting spam web pages through content analysis.
\newblock In {\em Proc.~of Int. Conf. on World Wide Web}, 2006.

\bibitem{Oard1998ImplicitFF}
D.~W. Oard and J.~Kim.
\newblock Implicit feedback for recommender systems.
\newblock In {\em Proceedings of the 1998 AAAI Workshop on Recommender
  Systems}, 1998.

\bibitem{pardos2014affective}
Z.~A. Pardos, R.~S. Baker, M.~O. San~Pedro, S.~M. Gowda, and S.~M. Gowda.
\newblock Affective states and state tests: Investigating how affect and
  engagement during the school year predict end-of-year learning outcomes.
\newblock {\em Journal of Learning Analytics}, 1(1), 2014.

\bibitem{scikit-learn}
F.~Pedregosa, G.~Varoquaux, A.~Gramfort, V.~Michel, B.~Thirion, O.~Grisel,
  M.~Blondel, P.~Prettenhofer, R.~Weiss, V.~Dubourg, J.~Vanderplas, A.~Passos,
  D.~Cournapeau, M.~Brucher, M.~Perrot, and E.~Duchesnay.
\newblock Scikit-learn: Machine learning in {P}ython.
\newblock {\em Journal of Machine Learning Research}, 12, 2011.

\bibitem{perezortiz2017practical}
M.~Perez-Ortiz and R.~K. Mantiuk.
\newblock A practical guide and software for analysing pairwise comparison
  experiments, 2017.

\bibitem{ramesh2014learning}
A.~Ramesh, D.~Goldwasser, B.~Huang, H.~Daume~III, and L.~Getoor.
\newblock Learning latent engagement patterns of students in online courses.
\newblock In {\em Proc.~of AAAI Conference on Artificial Intelligence}, 2014.

\bibitem{ShapiraTM06}
B.~Shapira, M.~Taieb{-}Maimon, and A.~Moskowitz.
\newblock Study of effectiveness of implicit indicators and their optimal
  combination for accurate inference of users interests.
\newblock {\em {JDIM}}, 4(3):169--174, 2006.

\bibitem{shawe2004kernel}
J.~Shawe-Taylor, N.~Cristianini, et~al.
\newblock {\em Kernel methods for pattern analysis}.
\newblock Cambridge university press, 2004.

\bibitem{slater2016semantic}
S.~Slater, R.~Baker, J.~Ocumpaugh, P.~Inventado, P.~Scupelli, and N.~Heffernan.
\newblock Semantic features of math problems: Relationships to student learning
  and engagement.
\newblock In {\em Proc.~of Int. Conf. on Educational Data Mining}, 2016.

\bibitem{taghipour2016neural}
K.~Taghipour and H.~T. Ng.
\newblock A neural approach to automated essay scoring.
\newblock In {\em Proc.~of Conf. on Empirical Methods in Natural Language
  Processing}, 2016.

\bibitem{Thurstone1927}
L.~L. Thurstone.
\newblock A law of comparative judgment.
\newblock {\em Psychological Review}, 34(4):273--286, 1927.

\bibitem{wiki_wang}
M.~Warncke-Wang, D.~Cosley, and J.~Riedl.
\newblock Tell me more: An actionable quality model for wikipedia.
\newblock In {\em Proc.~of Int. Symposium on Open Collaboration}, WikiSym '13,
  2013.

\bibitem{beyondviews}
S.~Wu, M.~Rizoiu, and L.~Xie.
\newblock Beyond views: Measuring and predicting engagement in online videos.
\newblock In {\em Proc.~of the Twelfth Int. Conf. on Web and Social Media},
  2018.

\bibitem{AES_dataset}
H.~Yannakoudakis, T.~Briscoe, and B.~Medlock.
\newblock A new dataset and method for automatically grading esol texts.
\newblock In {\em Proc.~of HLT}, 2011.

\bibitem{Zigoris2006}
P.~Zigoris and Y.~Zhang.
\newblock Bayesian adaptive user profiling with explicit \& implicit feedback.
\newblock In {\em Proceedings of the 15th ACM International Conference on
  Information and Knowledge Management}, CIKM '06, pages 397--404, New York,
  NY, USA, 2006. ACM.

\end{thebibliography}
\end{document}